\documentstyle[emulateapj]{article}
\slugcomment{submitted to ApJ, February 25, 2000}
\lefthead{BESKIN \& KUZNETSOVA}
\righthead{GRAD-SHAFRANOV EQUATION WITH ANISOTROPIC PRESSURE}

\begin{document}

\title{Grad-Shafranov equation with anisotropic pressure}

\author{Vasily S. Beskin}

\affil{P.N. Lebedev Physical Institute,
Russian Academy of Sciences,
Moscow, 117924, Russia; beskin@lpi.ru \\
DAEC, Observatory Paris -- Meudon, 92125, Meudon, France}

\vspace{.3cm}

\author{and}

\author{Inga V. Kuznetsova}
\affil{Moscow Institute of Physics and Technology, Dolgoprudnyi,
141700, Russia; kuzn@lpi.ru}

\begin{abstract}
The most general form of the nonrelativistic Grad-Shafranov equation
describing anisotropic pressure effects is formulated within the double
adiabatic approximation. It gives a possibility to analyze quantitatively
how the anisotropic pressure affects the 2D structure of the ideal
magnetohydrodynamical flows.
\end{abstract}


\section{Introduction}
The Grad-Shafranov (GS) equation for axisymmetric stationary ideal
magnetohydrodynamical (MHD) flows has been considered very intensively
within the last two decades both
for nonrelativistic  stellar (solar) wind including
outflow from young stellar objects
(Okamoto 1975; Heinemann \& Olbert 1978;
Blandford \& Payne 1982;
Heyvaerts \& Norman 1989; Sakurai 1990;
Pelletier \& Pudritz 1992; Shu et al 1994),
and for relativistic outflows from radio pulsars
(Ardavan 1979; Bogovalov 1992)
and 'central engine' in active galactic nuclei
(Phinney 1983; Begelman, Blandford \& Rees 1984; Mobarry \& Lovelace 1988;
Chiueh, Li \& Begelman 1991;
Beskin, Kuznetsova \& Rafikov 1998).
As is well known, in this approach the isotropic gas pressure
has been postulated (see e.g. Heyvaerts 1996 for introduction).

It is necessary to stress that, actually, thermal effects were considered
rather carefully only within the
given poloidal magnetic field approximation
(Weber \& Davis 1967;
Kennel, Fujimura \& Okamoto 1983;
Paatz \& Camenzind 1996),
or within the self-similar approach (Low \& Tsinganos 1986;
Contopoulos \& Lovelace 1994;
Sauty \& Tsinganos 1994).
As to the GS equation, its 2D solutions with nonzero pressure
were only obtained for pure hydrodynamical flows
(see Beskin 1997 for review) when the isotropic pressure model is correct.

On the other hand, for real astrophysical flows where the effects of
finite pressure may play an important role, the isotropic pressure model
does not seem to be sufficient. For example, in the solar wind where the
free path $l_{\nu}$ exceeds the Larmor radius $r_{H}$
by many order of magnitude,
so that $l_{\nu}/r_{H} \sim 10^9$ in the vicinity of the Earth,
the effect of the anisotropic pressure may be important.

Of course, it is necessary to remember that the ideal MHD approximation
(including the double adiabatic approach under consideration)
is not good for real astrophysical objects,
and so the dissipative processes can play an important role
(Leer \& Axford 1972; Hu, Esser \& Habbal 1997).
Moreover, in the solar wind the stochastical component of the
magnetic field is of the order of the regular one.
Such fluctuations drastically change the velocity distribution
function. As a result, the pressure anisotropy
$P_{\parallel}/P_{\perp} \approx 1.2$
is not actually large (see e.g. Marsch et al 1982).
May be, for this reason the effects of the anisotropic pressure on the
outflow structure within the GS approach were analyzed in a small number
of papers (Ass\'eo \& Beaufils 1983; Tsikarishvili, Rogava \& Tsikauri 1995),
the GS equation itself being formulated only for the planar
statical configurations with zero longitudinal electric current
(N\"otzel, Schindler \& Birn 1985).

In this paper we give the general equations
describing the axisymmetric stationary ideal MHD flows with
anisotropic pressure. Then, we shall present the results of
our preliminary analysis of the effects of the anisotropic pressure
(such an analysis has not been produced previously either).

\section{Basic Equations}

Let us consider a stationary axisymmetric (and nonrelativistic) MHD flow.
Then the magnetic field can be written in the form
\begin{equation}
 {\bf B} =\frac{{\bf \nabla}\Psi \times {\bf e}_{\hat \varphi}}{2\pi\varpi}
  -\frac{2I}{\varpi c}{\bf e}_{\hat \varphi}.
    \label{frst}
     \end{equation}
Here $\Psi(r,\theta)$ is the magnetic flux, $\varpi = r\sin\theta$, and
$I(r,\theta)$ is the total electric current flowing inside
the region $\Psi<\Psi(r,\theta)$. Assuming that the magnetosphere contains
enough plasma to screen the longitudinal electric field $E_{\parallel}$, one
can write down
\begin{equation}
 {\bf E}=-\frac{\Omega_{\rm F}}{2\pi c}{\bf \nabla}\Psi,
  \label{3}
   \end{equation}
the 'field angular velocity'
$\Omega_{\rm F}=\Omega_{\rm F}(\Psi)$ being constant on
the magnetic surfaces $\Psi(r,\theta)=$ const.
Finally, the continuity equation
${\bf \nabla}\cdot(n{\bf u})=0$ together with the freezing-in condition
${\bf E} + {\bf v}\times{\bf B}/c = 0$ and the Maxwell equation
${\bf \nabla} \times {\bf E} = 0$ result in
\begin{equation}
 {\bf v}=\frac{\eta}{\rho}{\bf B}+\Omega_{\rm F}
  \varpi{\bf e}_{\hat\varphi},
  \label{4a}
   \end{equation}
where $\rho$ is the mass density and $\eta = \eta(\Psi)$ is the
particle-to-magnetic flux ratio. Clearly, these expressions do not depend on the
pressure anisotropy.

The next step is the transformation of the Euler equation
\begin{equation}
\rho({\bf v}\cdot{\bf \nabla}){\bf v} = -{\bf \nabla}_{\beta}P_{\alpha\beta}
+\frac{1}{4\pi}[{\bf \nabla}\times{\bf B}]\times{\bf B} +\rho{\bf g}.
\label{Eu}
\end{equation}
According to Chew, Goldberger \& Low (1956),
for collisionless plasma with an anisotropic pressure
\begin{equation}
P_{\alpha\beta} = P_s\delta_{\alpha\beta}
+(P_n - P_s)B_{\alpha}B_{\beta}/B^2
\end{equation}
it is possible to introduce two extra invariants on the
magnetic surfaces $\Psi(r,\theta) =$ const
(the so-called double adiabatic approximation
resulting from the concervation of two adiabatic invariants)
\begin{eqnarray}
s_1(\Psi) & = & \frac{P_n B^2}{\rho^3},
\label{s1} \\
s_2(\Psi) & = & \frac{P_s}{\rho B}.
\label{s2}
\end{eqnarray}
They correspond to polytropic equations of state with $\Gamma_{\parallel} = 3$
and $\Gamma_{\perp} = 1$.
As a result, integrating ${\bf v}$ and $\varphi$ components of
equation (\ref{Eu}), one can obtain the known expressions for
the energy and the angular momentum
(Ass\'eo \& Beaufils 1983; Tsikarishvili, Rogava \& Tsikauri 1995):
\begin{eqnarray}
E(\Psi) & = & \frac{v^2}{2}
+ \frac{P_s}{\rho} +\frac{3}{2}\,\frac{P_n}{\rho}
+ \frac{\Omega_{\rm F}I}{2\pi c\eta}\left(1 - \beta_{\rm a}\right)
-\frac{GM}{r},
\label{B} \\
L(\Psi) & = & v_{\varphi}r\sin\theta
+ \frac{I}{2\pi c\eta}\left(1 - \beta_{\rm a}\right).
\end{eqnarray}
Here
\begin{equation}
\beta_{\rm a} = 4\pi\frac{P_n - P_s}{B^2}
\end{equation}
is the anisotropic pressure parameter.
Together with $\Omega_{\rm F}(\Psi)$ and $\eta(\Psi)$
these six invariants determine all the characteristics of a flow.
In particular,
\begin{eqnarray}
\frac{I}{2\pi}& = & c\eta
\frac{L - \Omega_{\rm F}\varpi^2}{1-M^2-\beta_{\rm a}},
\label{I} \\
v_{\varphi} & = & \frac{1}{\varpi}\cdot
\frac{\Omega_{\rm F}\varpi^2(1-\beta_{\rm a}) - M^2 L}
{1-M^2-\beta_{\rm a}},
\label{vphi}
\end{eqnarray}
where $M^2 = 4\pi\eta^2/\rho$
is the Alfv\'enic Mach number.
Thus, in the anisotropic case the Alfv\'enic singularity
has the form
\begin{equation}
A =  1 - M^2 -\beta_{\rm a} = 0.
\label{A}
\end{equation}

It is necessary to stress that (\ref{I}) gives an implicit expression
for the current $I$ because $I$ through $B_{\varphi}^2$ is contained
in the rhs of this equation as well. Nevertheless, the standard
procedure of the determination of flow parameters in a given
poloidal magnetic field ${\bf B}_{\rm p}$
remains the same. The Bernoulli equation (\ref{B}), which can be rewritten
in the form
\begin{eqnarray}
\frac{M^4}{64\pi^4\eta^2}({\bf \nabla}\Psi)^2
= 2\varpi^{2}\left(E-\frac{P_s}{\rho} -
\frac{3}{2}\frac{P_n}{\rho} + \frac{GM}{r}\right) \nonumber \\
-\frac{[\Omega_{\rm F}\varpi^2(1-\beta_{\rm a})-LM^2]^2}
{(1-M^2-\beta_{\rm a})^2}
\nonumber \\
-2\varpi^2\Omega_{\rm F}(1-\beta_{\rm a})\frac{L-\Omega_{\rm F}\varpi^2}
{1-M^2-\beta_{\rm a}},
\label{b1a}
\end{eqnarray}
taken together with (\ref{I}) and with the definitions (\ref{s1}), (\ref{s2})
determines implicitly
the Mach number $M^2$ and $\beta_{\rm a}$
as functions of $\Psi$ (or ${\bf B}_{\rm p}$)
and all the six invariants
\begin{eqnarray}
M^2 & = &
         M^2[({\bf \nabla}\Psi)^2, E, L, \Omega_{F}, \eta, s_1, s_2],
\label{mm} \\
\beta_{\rm a} & = &
         \beta_{\rm a}[({\bf \nabla}\Psi)^2, E, L, \Omega_{F}, \eta, s_1, s_2].
\label{bb}
\end{eqnarray}
It means that all the other parameters ($I$, $v_{\varphi}$, etc.)
can be found from given values of ${\bf B}_{\rm p}$ and of all the
six invariants as well.
This fact is fully analogous to spherically symmetric flows
(Bondi 1952; Parker 1958) where the flow structure is known
and the algebraic expressions for the
integrals of motion are enough to determine all the characteristics of a flow.

On the other hand, if the flow is not spherically symmetric,
the poloidal magnetic field is to be
determined from the ${\bf \nabla}\Psi$ component of equation (\ref{Eu}).
After numerous but elementary calculations it can be presented
in the following compact form
\begin{eqnarray}
  {\bf \nabla}_{k}\left[\frac{1}{\varpi^{2}}
   \left(1-M^2-\beta_{\rm a}\right)
    {\bf \nabla}^{k}\Psi\right]
     +\frac{64\pi^{4}}{\varpi^{2}}\frac{1}{2M^{2}}\frac{\partial}
      {\partial\Psi}\left(\frac{G}{A}\right) \nonumber \\
       -8\pi^3P_n\frac{1}{s_1}\frac{{\rm d}s_1}{{\rm d}\Psi}
        -16\pi^3P_s\frac{1}{s_2}\frac{{\rm d}s_2}{{\rm d}\Psi} = 0.
         \label{GS}
\end{eqnarray}
Here
\[
  \left(\frac{G}{A}\right)
   =2\varpi^{2}\eta^2\left(E-\frac{P_s}{\rho}
    -\frac{3}{2}\frac{P_n}{\rho} + \frac{GM}{r}\right) \nonumber \]
\begin{equation}  \quad
  +\eta^2\frac{\varpi^4\Omega_{\rm F}^2(1-\beta_{\rm a})
     -2\varpi^2\Omega_{\rm F}L(1-\beta_{\rm a})
     +M^{2}L^{2}}{1-M^2 - \beta_{\rm a}},
       \label{20b}
         \end{equation}
${\bf \nabla}_k$ is a covariant derivative,
and the operator $\partial/\partial\Psi$ acts on the invariants
$E(\Psi)$, $L(\Psi)$, $\Omega_{\rm F}(\Psi)$, and $\eta(\Psi)$ only.
According to (\ref{s1})--(\ref{s2}) and (\ref{mm})--(\ref{bb}),
this trans-field (stream, generalized GS) equation contains
the unknown function $\Psi(r,\theta)$ and the six
invariants only. As to the GS equation itself
describing statical configurations, it can be obtained from (\ref{GS})
in the limit
$\Omega_{\rm F} \rightarrow 0$, $\eta \rightarrow 0$
(and hence $M^2 \rightarrow 0$), but $\eta L \rightarrow $ const.
It gives for gravity-free case
\begin{eqnarray}
  {\bf \nabla}_{k}\left[\frac{1}{\varpi^{2}}
   \left(1-\beta_{\rm a}\right)
    {\bf \nabla}^{k}\Psi\right]
     +\frac{16\pi^{2}}{\varpi^{2}}(1-\beta_{\rm a})
      I\frac{{\rm d}I}{{\rm d} \Psi}\nonumber \\
       +16\pi^3 \rho \frac{{\rm d}}{{\rm d}\Psi}
        \left(\frac{P_s}{\rho} + \frac{3}{2}\frac{P_n}{\rho}\right)
         \nonumber \\
         -8\pi^3P_n\frac{1}{s_1}\frac{{\rm d}s_1}{{\rm d}\Psi}
          -16\pi^3P_s\frac{1}{s_2}\frac{{\rm d}s_2}{{\rm d}\Psi}=0.
          \label{GS0}
\end{eqnarray}

Relations (\ref{B})--(\ref{vphi}) remain true even
for a more general bounded anisotropy model (Denton et al 1994)
\begin{eqnarray}
({\bf v}\cdot{\bf \nabla})\left(\frac{P_n B^2}{\rho^3}\right) & = &
2{\cal P}_t\left(\frac{B^2}{\rho^3}\right),
\label{s1m} \\
({\bf v}\cdot{\bf \nabla})\left(\frac{P_s}{\rho B}\right) & = &
-{\cal P}_t\left(\frac{1}{\rho B}\right),
\label{s2m}
\end{eqnarray}
where ${\cal P}_t$ is a nondissipative energy exchange term.
In this case we have the concervation of the 'total entropy' $S = S(\Psi)$
\begin{equation}
({\bf v}\cdot{\bf \nabla})S
= \frac{1}{2} \frac{\rho^3}{B^2}({\bf v}\cdot{\bf \nabla})s_1
+ \rho B ({\bf v}\cdot{\bf \nabla})s_2 = 0,
\end{equation}
and so the generalized GS equation (\ref{GS}) has a form
\begin{eqnarray}
  {\bf \nabla}_{k}\left[\frac{1}{\varpi^{2}}(1 - M^2 - \beta_{\rm a})
   {\bf \nabla}^{k}\Psi\right]
    +\frac{32\pi^{4}}{\varpi^{2}M^2}\frac{\partial}
      {\partial\Psi}\left(\frac{G}{A}\right) \nonumber \\
       -16 \pi^3 \frac{{\rm d}S}{{\rm d}\Psi} = 0.
         \label{GSm}
\end{eqnarray}
Introducing the effective pressure $P_{\rm eff} = P_{\rm eff}(\Psi)$ as
\begin{equation}
{\rm d}P_{\rm eff} = \rho \, {\rm d}\left(\frac{P_s}{\rho}
+ \frac{3}{2}\frac{P_n}{\rho}\right) - {\rm d}S,
\end{equation}
which is equivalent to the ordinary thermodynamic relation
${\rm d}P = \rho {\rm d}w - \rho T{\rm d}s$,
one can obtain instead of (\ref{GS0})
\begin{eqnarray}
  {\bf \nabla}_{k}\left[\frac{1}{\varpi^{2}}
   \left(1-\beta_{\rm a}\right)
    {\bf \nabla}^{k}\Psi\right]
     +\frac{16\pi^{2}}{\varpi^{2}}(1-\beta_{\rm a})
      I\frac{{\rm d}I}{{\rm d} \Psi} \nonumber \\
        + 16\pi^3\frac{{\rm d}P_{\rm eff}}{{\rm d} \Psi} = 0.
          \label{GS0m}
\end{eqnarray}
For $I = 0$ it has the form found by
N\"otzel, Schindler \& Birn (1985) for planar geometry.

Equation (\ref{GS}) is a partial equation of the mixed type.
Using the implicit relations resulting in (\ref{mm})--(\ref{bb}),
one can find for the second-order operator in (\ref{GS})
\begin{equation}
  A\left[{\bf \nabla}_{k}\left(\frac{1}{\varpi^{2}}
   {\bf \nabla}^{k}\Psi\right) +
    \frac{({\bf \nabla}^{i}\Psi)({\bf \nabla}^{k}\Psi){\bf \nabla}_{i}
     {\bf \nabla}_{k}\Psi}{\varpi^{2}({\bf \nabla}\Psi)^{2}D}\right] + \dots =
0,
\label{op}
\end{equation}
where
\begin{equation}
D = \frac{n}{d},
\label{D}
\end{equation}
and
\begin{eqnarray}
n = \left(1-M^2-\beta_{\rm a}
+4\pi\frac{4P_n-P_s}{B^2}\frac{B_{\varphi}^2}{B^2}\right)
\left(1 -3 \frac{P_n}{\rho v_{\rm p}^2}\right)
\nonumber \\
+\frac{B_{\varphi}^2}{B_{\rm p}^2}
\left(1 -\beta_{\rm a} -
4\pi\frac{3P_n-P_s}{B^2}
+4\pi\frac{4P_n-P_s}{B^2}\frac{B_{\varphi}^2}{B^2}
\right)
\nonumber \\
-\left(M^2 - 4\pi\frac{3P_n-P_s}{B^2}\right)
\frac{3P_n-P_s}{\rho v_{\rm p}^2}\frac{B_{\varphi}^2}{B^2},
\end{eqnarray}
\begin{eqnarray}
d =\left(1-3\frac{P_n}{\rho v_{\rm p}^2}\frac{B_{\rm p}^2}{B^2}\right)
     \left(M^2 + 4\pi \frac{P_n}{B^2}\right)
    \nonumber \\
     + \frac{P_s}{\rho v_{\rm p}^2}\frac{B_{\rm p}^2}{B^2}
     \left(M^2 - 4\pi\frac{3P_n-P_s}{B^2}\right).
\end{eqnarray}
The  form (\ref{op}) of equation (\ref{GS})
coincides with the one found for an isotropic pressure
(see e.g. Sakurai 1990; Beskin 1997). Hence, one can conclude that
the trans-field equation is elliptical for $D > 0$ and $D < -1$,
and hyperbolic for $-1 < D < 0$. It means that the trans-field
equation changes from elliptical to hyperbolic on the surfaces $D = 0$
(fast and slow magnetosonic singularities) and $D = -1$ (cusp singularity).

It is necessary to stress that, as in the isotropic case, $D$
can be presented as $D = d_1A + d_2B_{\varphi}^2/B^2$, so that
$D \propto A$ for $B_{\varphi}=0$. This property reflects the fact
that for ${\bf k} \parallel {\bf B}$ fast or slow magnetosonic velocity
coincides with the Alfv\'enic one.
In particular, for a nonrotating flow ($\Omega_{\rm F} = 0$, $L = 0$),
when, according to (\ref{I}), $B_{\varphi} =0$,
we have
\begin{equation}
n = (1-M^2-\beta_{\rm a})\left(1-3\frac{P_n}{\rho v_{\rm p}^2}\right).
\end{equation}
Using now the definitions (\ref{A}) and (\ref{D}), one can obtain
for sonic, Alfv\'enic, and cusp velocities in anisotropic media
\begin{eqnarray}
a_s^2 & = & 3\frac{P_n}{\rho},
\label{as} \\
v_{\rm A}^2 & = & \frac{B^2}{4\pi\rho} -\frac{P_n-P_s}{\rho}, \\
v_{\rm c}^2 & = & \frac{\displaystyle{3\frac{P_n}{\rho}\frac{B^2}{4\pi\rho}
+6\frac{P_nP_s}{\rho^2} - \frac{P_s^2}{\rho^2}}}
{\displaystyle{\frac{B^2}{4\pi\rho} +2\frac{P_s}{\rho}}}
\label{vc}.
\end{eqnarray}
As we see, the velocities (\ref{as})--(\ref{vc}) coincide
exactly with the characteristic velocities propagating
in anisotropic media (Clemmow \& Dougherty 1969).
Recall that the condition $v_{\rm A}^2 < 0$ corresponds to the firehose
instability, and the condition $v_{\rm c}^2 < 0$ -- to the mirror instability.

\section{Two examples}

\subsection{Free outflow}

As a first example, let us consider a free (i.e. without gravity effects)
outflow along a strong monopole magnetic field from the surface of a
nonrotating sphere (with radius $r_0$, magnetic field $B_0$,
pressure $P_0 \ll B_0^2/4\pi$, mass density $\rho_0$,
and radial velocity $v_0$). It is natural to assume that here
$P_s(r_0) = P_n(r_0) = P_0$. As a result,
one can find for the radial velocity
\begin{eqnarray}
\frac{v^2}{v_0^2} = \frac{1}{2} +
\frac{5}{2}p\left(1-\frac{2}{5}\frac{r_0^2}{r^2}\right)  \nonumber \\
+\frac{1}{2}
\sqrt{\left[1+5p\left(1-\frac{2}{5}\frac{r_0^2}{r^2}\right)\right]^2 - 12p},
\label{vf}
\end{eqnarray}
where
\begin{equation}
p = \frac{P_0}{\rho_0v_0^2}.
\end{equation}
It gives
\begin{equation}
r\frac{{\rm d}(v^2/v_0^2)}{{\rm d}r} = \frac{N_{v}}{D},
\label{dvdr}
\end{equation}
where $N_{v} = 4pr_0^2/r^2$, and
\begin{equation}
D = 1 - 3\frac{P_n}{\rho v^2},
\label{Df}
\end{equation}
which corresponds to the longitudinal sonic singularity in (\ref{D}).
On the other hand, the same as for isotropic media, for a nonrotating wind
the magnetosonic surface coincides with the Alfv\'enic one,
the singularity for $1-M^2-\beta_{\rm a} = 0$ being absent.

Comparing now expression (\ref{vf}) for $v(r)$
with the appropriate one corresponding to an outflow
with an isotropic pressure for the polytropic equation of state with
$\Gamma = 3$ and the same energy $E$,
one can conclude that the difference only takes place
for small enough pressure $P_0 \ll \rho_0v_0^2$ when
\begin{eqnarray}
\frac{v_{\infty}^{({\rm iso})}}{v_0} & = &
         1+\frac{1}{6}\left(\frac{2E}{v_0^2} - 1\right), \\
\frac{v_{\infty}^{({\rm aniso})}}{v_0} & = &
  1+\frac{1}{5}\left(\frac{2E}{v_0^2} - 1\right).
\end{eqnarray}
For $P_0 \gg \rho_0v_0^2$ the limiting velocity does not depend on the pressure
model: $v_{\infty}^2 \approx 2E$. Thus, the difference takes place for
$r \sim r_0$ only.

It is also interesting to determine how the anisotropic pressure
affects the position of the Alfv\'enic surface $r_{\rm cr}$
coinciding for $\Omega_{\rm F} = 0$ with the fast magnetosonic one. As
\begin{equation}
\frac{\beta_{\rm a}}{M^2} \approx p\left(\frac{v_0}{v}\right)^4,
\label{bp}
\end{equation}
it is easily check that $\beta_{\rm a}/M^2 \ll 1$ for both $p \ll 1$ and
$p \gg 1$, so that for $P_0 \ll \rho_0v_0^2$ and
$P_0 \gg \rho_0v_0^2$ the position of the Alfv\'enic surface
\begin{equation}
r_{\rm cr}^2 \approx r_{\rm 0}^2\frac{B_0^2}{4\pi\rho_0v_0^2}
\,\frac{v_0}{v_{\infty}}
\end{equation}
does not depend on the pressure anisotropy. Finally, as
\begin{equation}
\frac{P_n}{P_s} = \frac{v_0^2}{v^2}\frac{r^2}{r_0^2},
\label{pp}
\end{equation}
the anisotropy of the pressure tensor increases with distance
from the central object.
As has already been stressed, it can be realized only if
dissipative processes (heat conductivity, plasma instability, etc.)
play no role.
Relations (\ref{bp}) and (\ref{pp}) also mean that the conditions for
the firehose and mirror instabilities
are not satisfied up to the Alfv\'enic surface.
On the other hand, the firehose instability takes place
in the supersonic region
for $r > r_{\rm fh} \gg r_{\rm cr}$, where
\begin{equation}
r_{\rm fh}^2 \approx r_{\rm cr}^2\frac{1}{p}
\left(\frac{v_\infty}{v_0}\right)^4.
\end{equation}

\subsection{Parker outflow}

As a next example we consider an outflow from the gravity center of mass
$M$ in a monopole magnetic field (Parker ejection).
Analyzing the energy equation (\ref{B}), we have
\begin{eqnarray}
\frac{v^2}{v_0^2} = \frac{1}{2} + \frac{5}{2}p - p\frac{r_0^2}{r^2}
+\frac{GM}{v_0^2}\frac{r_0 - r}{rr_0}
\nonumber \\
+ \frac{1}{2}
\sqrt{\left(1 + 5p - 2p\frac{r_0^2}{r^2}
+2\frac{GM}{v_0^2}\frac{r_0 - r}{rr_0}\right)^2 - 12p}.
\label{P}
\end{eqnarray}
As a result, now
$N_v  =  4pr_0^2/r^2 - 2GM/v_0^2r$,
and $D$ is determined by (\ref{Df}) again.
Hence, for the position of the sonic surface we have
\begin{equation}
r_s = 2r_0\frac{P_0r_0}{GM\rho_0}.
\end{equation}
On the other hand, one can easily check that equation (\ref{P})
has a real solution for $r_s \leq r_0$ only.
This result is in agreement with a well-known fact that
no transonic outflow is possible for a gas with a polytropic index $\Gamma = 3$.
Simultaneously, for Parker ejection the conditions of the
firehose and mirror instabilities
up to Alfv\'enic surface are not satisfied either.

\section{Conclusion}

Thus, we have generalized the GS equation for
the flows with anisotropic pressure.
It allows a quantitative analysis of the effect of
anisotropic pressure upon the 2D outflow structure
within ideal double adiabatic approximation.
As has been shown on simple examples, the pressure anisotropy itself
may be very large, but not in all cases does it result in a large
disturbance of the flow parameters.
Clearly, the additional analysis (say, within the Weber-Davis model)
is necessary to clarify all the details of the flow with
anisotropic pressure.

This work was supported by the Russian Foundation for Basic Research
(project No. 99-02-17184). VSB thanks
A.A.~Samsonov for stimulating discussions and
Observatory Paris -- Meudon for hospitality.
IVK thanks the International Soros Educational
Program (Grant a99-941) and Landau Foundation for financial support.


\begin{thebibliography}{}


  \bibitem[]{}Ardavan, H. 1979, MNRAS, 189, 397
  \bibitem[]{}Ass\'eo, E, \& Beaufils, D.  1983, Ap\&SS, 89, 133
  \bibitem[]{}Begelman, M.C., Blandford, R.D., \& Rees, M.J. 1984,
Rev. Mod. Phys., 56, 255
  \bibitem[]{}Beskin, V.S. 1997, Physics Uspekhi, 40(7), 659
  \bibitem[]{}Beskin, V.S., Kuznetsova, I.V., \& Rafikov, R.R. 1998, MNRAS,
299, 341
  \bibitem[]{}Blandford, R.D., \& Payne, D.G. 1982, MNRAS, 199, 883
  \bibitem[]{}Bogovalov, S.V. 1992, Sov. Astron. Lett., 18, 337
  \bibitem[]{}Bondi, H. 1952, MNRAS, 112, 195
  \bibitem[]{}Chew, G.F., Goldberger, M.L., \& Low, F.E. 1956,
Proc. Roy. Soc. London A, 236, 112
  \bibitem[]{}Chiueh, T., Li, Zh.-Y., \& Begelman, M.C. 1991, ApJ, 377, 462
  \bibitem[]{}Clemmow, P.C., \& Dougherty, J.P. 1969, Electrodynamics of
Particles and Plasmas, (Reading: Addisson-Wesley Publ. C.), 385
  \bibitem[]{}Contopoulos, J., \& Lovelace, R.V.E. 1994, ApJ, 429, 139
  \bibitem[]{}Denton, R.E., Anderson, B.J., Gary, S.P., \& Fuselier, S.A.
1994, J. Geophys. Res., 99, 11225
  \bibitem[]{}Heinemann, M., \& Olbert, S. 1978, J. Geophys. Res., 83, 2457
  \bibitem[]{}Heyvaerts, J., 1996, in Plasma Astrophysics,
ed. C. Chiuderi \& G. Einaudi
(Berlin: Springer), 31
  \bibitem[]{}Heyvaerts, J., \& Norman, C. 1989, ApJ, 347, 1055
  \bibitem[]{}Hu, Y.Q., Esser, R., \& Habbal, S.R. 1997, J. Geophys. Res.,
102, 14661
  \bibitem[]{}Kennel, C.F., Fujimura, F.S., \& Okamoto, I. 1983, Geophys.
Astrophys. Fluid Dynamics, 26, 147
  \bibitem[]{}Leer, L., \& Axford, W.I. 1972, Sol. Phys., 23, 238
  \bibitem[]{}Low, B.C., \& Tsinganos, K. 1986, ApJ, 302, 163
  \bibitem[]{}Marsch, E., et al. 1982, J. Geophys. Res., 87, 52
  \bibitem[]{}Mobarry, C.M., \& Lovelace, R.V.E. 1986, ApJ, 309, 455
  \bibitem[]{}N\"otzel, A., Schindler, K., \& Birn, J. 1985, J. Geophys. Res.,
90, 8293
  \bibitem[]{}Okamoto, I. 1975, MNRAS, 173, 357
  \bibitem[]{}Paatz, G., \& Camenzind, M. 1996, A\&A, 313, 591
  \bibitem[]{}Parker, E.N. 1958, ApJ, 128, 664
  \bibitem[]{}Pelletier, G., \& Pudritz, R.E. 1992, ApJ, 394, 117
  \bibitem[]{}Phinney, E. 1983, in Astrophysical Jets, ed.
A.Ferrary \& A.G.Pacholczyk, (Dodrecht: Reidel), 201
  \bibitem[]{}Sakurai, T. 1990, Computer Phys. Rep., 12, 247
  \bibitem[]{}Sauty, C., \& Tsinganos, K. 1994, A\&A, 287, 893
  \bibitem[]{}Shu, F.H., et al. 1994, ApJ, 429, 797
  \bibitem[]{}Tsikarishvili, E.G., Rogava, A.D., \& Tsikauri, D.G. 1995,
ApJ, 439, 822
  \bibitem[]{}Weber, E.J., \& Davis, L. Jr 1967, ApJ, 148, 217

\end{thebibliography}
\end{document}